\documentclass[aps,prd,twocolumn,groupedaddress]{revtex4}
\usepackage{amsmath, amsfonts}
\usepackage[final]{graphicx}

\begin{document}

\title{Heavy Charged Leptons in 6-dimensional Randall-Sundrum Model}

\author{Erin De Pree}\email[]{ekdepree@smcm.edu}
\author{Dietrich Kiesewetter}\email[]{dkiesewetter@gmail.com}
\affiliation{Department of Physics, St. Mary's College of Maryland, St. Mary's City, MD 20686, USA}

\date{\today}

\begin{abstract}
	
	We consider the effects of adding a fourth generation to a model with two compactified extra dimensions -- one warped and the other flat.  The extra dimensions lead to towers of Kaluza-Klein $Z$ bosons.  Interestingly, the higher KK-$Z$ modes depending on the flat dimension contribute more than the warped dimension.
	
	These KK-$Z$ modes enhance the cross-section of fourth generation charged leptons .  The most promising signature is the heavy charged lepton decaying to a heavy neutrino and $W$ boson.

\end{abstract}



\maketitle

\section{Introduction}
	
	The Randall-Sundrum (RS) model \cite{RS} rather nicely explains the hierarchy problem.  The metric for the RS model compactified on $S_1$ is
	\begin{equation}
		ds^2 = e^{-2\sigma} \eta_{\mu\nu} dx^\mu dx^\nu - r_c^2 d\phi^2
	\end{equation}
where $\eta_{\mu\nu}$ is the Minkowski flat metric, $\mu, \nu = 0, 1, 2, 3$ and $\phi \in [-\pi, \pi]$ is the warped extra dimension.  The RS warp factor is $\sigma (\phi ) = kr_c | \phi |$ where $k$ is the scale of curvature and $r_c$ is the radius of compactification.  In order to ensure that the TeV brane is scaled to about a TeV from the IR (or Planck) brane, $kr_c \sim 11$, these issues (as well as many others) are throughly discussed in reference \cite{kR}.
	
	Allowing fermions to live in the AdS bulk reduces the fermion mass hierarchy problem to a simple geography problem \cite{fermions_in_bulk}.  The mass of the fermion depends on its coupling to the Higgs (which resides on the TeV brane) which is determined by the fermion's position in the bulk.  Fermions far away from the TeV brane couple weakly with the Higgs and are thus less massive, while fermions residing close to the TeV brane couple strongly to the Higgs and, thus, are more massive.  
	
	Recently the RS model has been extended to add an extra flat dimension \cite{RS6}.  We explore the consequences of adding a fourth generation to this six dimensional model.

	A fourth generation of fermions has been considered off and on for many years \cite{4gen}.  Although a fourth generation and warped extra dimensions have been studied \cite{4genRS}, in this paper we consider a fourth generation and warped extra dimension with an additional flat dimension.

\section{Expanding to 6 Dimensions}
	
	A recent extension \cite{RS6} of the Randall-Sundrum model \cite{RS}, adds an extra flat dimension -- this model is named RS6 for convenience.  Reference \cite{RS6} considers the extra dimension compactified on $S^1$ and $S^1/Z_2$, however we will consider only the $S^1$ compactification case.  The $S^1/Z_2$ case is closely related, the coupling constant $g_{nl;c}$ simply increases by a factor of $\sqrt{2}$.
	
	In this 6 dimensional space, we have the regular 4 dimensions, the warped dimension $x_4 = r_c \phi$, and a sixth dimension $x_5 = R\theta$.  The metric for this model is 
	\begin{equation}
		ds^2 = e^{-2\sigma} \eta_{\mu\nu} dx^\mu dx^\nu - r_c^2 d\phi^2 - R^2 d\theta^2
		\label{RS6 metric}
	\end{equation}
	where $\sigma (\phi) = kr_c | \phi |$ is the typical RS warp function, $k$ is the scale of curvature, $r_c$ is the radius of compactification of the $\text{AdS}_5$ slice, and $\phi \in [-\pi, \pi]$ and is $Z_2$ orbifolded.  As usual, $kr_c \sim 11$.  Finally, $R$ is the radius of $S^1$ and $\theta \in [0,2\pi]$ and $kR \sim 1$.  The gravitational sector of this model has already been studied \cite{RS6_grav}.
	
	Davoudiasl and Rizzo \cite{RS6} explored the gauge bosons in RS6 and compared the results with the original RS model.  We expand on this model by adding a fourth generation of fermions.

\section{Fourth Generation}
	
	We first introduce a sequential fourth generation; in this paper we will focus on charged heavy leptons.  These effects will will not have any noticeable impact on the fourth generation quarks because of the large tree level diagrams.  
	
	The $c$ value describes where in the warped dimension the fermion is located.  As the fermion position moves closer to the TeV brane (where the Higgs is located), $c$ approaches negative infinity.  Because the heavy lepton mass is greater than $100$ GeV, it will be closer to the TeV brane than the bottom quark -- almost as close as the right-handed top quark.  This allows us to use the useful approximations to the fermion wavefunctions and coupling constants.  Notice the coupling constants $g_{nl;c}$ of the KK gauge bosons and SM fermions in RS6 \cite{RS6} is 
	\begin{equation}
		g_{nl;c} = \sqrt{2} \int_{-\pi}^\pi f_0^2 \chi^{(n,l)} (\phi) d\phi
	\end{equation}
	where $f_0$ is the bulk profile of the fermions
	\begin{equation}
		f_0 = e^{(1/2 - c) \sigma(\phi)} \sqrt{ \frac{ kr_c (1/2-c) }{e^{kr_c\pi(1-2c)} - 1} }
	\end{equation}
	and $\chi^{(n,l)}$ is the KK gauge boson wavefunction
	\begin{equation}
		\chi^{(n,l)} (\phi) = \frac{ e^\sigma }{ N_{nl} } \left ( J_\nu (z_{nl} ) + \alpha_{nl} Y (z_{nl}  ) \right )
	\end{equation}
	which is normalized by
	\begin{equation}
		N_{nl} = \frac{e^{kr_c\pi}}{x_{nl} \sqrt{ kr_c } } \sqrt{ \left . \left [ \zeta_\nu^2 (z_{nl} )  \left ( z_{nl}^2 -\nu_l^2 + 1 \right ) \right ] \right |_{\phi=0}^{\phi=\pi} }
	\end{equation}
	(note that the normalization is corrected from reference \cite{RS6} which contained a typographical error in the paper, but still agrees with the results and figures, this correction corresponds to another typographical error in reference \cite{cRS} and allows us to reproduce the graphs of $g_{nl;c}$)
	\begin{eqnarray}
		\zeta_{nl}(z_{nl}) &=& J_\nu (z_{nl}) + \alpha_{nl} Y_\nu (z_{nl}) \\
		\alpha_{nl} &=& - \left. \frac{J_\nu (z_{nl}) + z_{nl} J_\nu^\prime (z_{nl}) }{Y_\nu (z_{nl}) + z_{nl} Y_\nu^\prime (z_{nl}) } \right |_{\phi=0} \\
		z_{nl} (\phi) &=& x_{nl} e^{\sigma-kr_c\pi} \\
		\sigma (\phi) &=& kr_c \left| \phi \right| \\
		\nu_l &=& \sqrt{ 1 + \left( \frac{ l }{ kR } \right)^2 }
	\end{eqnarray}
	and $x_{nl}$ is the $n^\text{th}$ root of 
	\begin{equation}
		J_\nu (x_{nl}) + x_{nl} J_\nu^\prime (x_{nl}) = 0
	\end{equation}

	\begin{figure}
		\resizebox{0.9\columnwidth}{!}{ \includegraphics{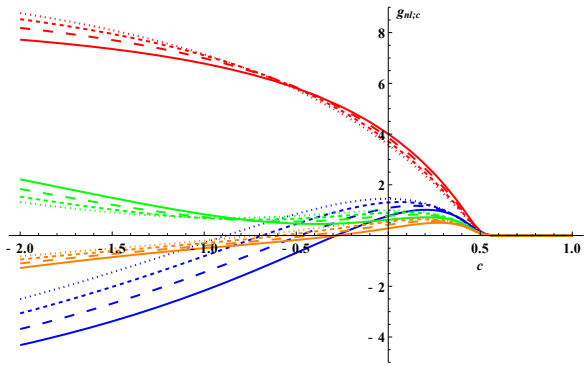} }
		\caption{The coupling $g_{nl;c}$ with the excited KK-Z bosons for $kR = 1$ and $kr_c=11$.  The red lines are for the 
			$n=1$ KK-gauge bosons, the blue lines for $n=2$, green for $n=3$, and orange for $n=4$.  For each excitation 
			in the warped extra dimension, there are also possible excitations in the flat dimension: the solid lines are for 
			$l=1$, the large dashes for the $l=2$, small dashes for $l=3$ and dotted for $l=4$.  Remember that $c>0.5$ 
			corresponds to light fermions and heavy fermions have $c<0.5$.}
		\label{fig.gnlc}
	\end{figure}
	
	Figure \ref{fig.gnlc} also indicates an increasingly large dependence on higher $Z^{(n,l)}$ bosons as $c$ decreases -- which is equivalent to the mass increasing.  Interestingly, for the $n=1$ excitations, the coupling constant $g_{1l;c}$ increase as $l$ increases.  This is opposite to what we expected and only occurs the $n=1$ values, not for higher $n$ values.  
	
	Note that in fig.\ \ref{fig.gnlc} the $n=1$ KK $Z$ bosons couple much more strongly than larger $n$ KK modes.  So the most impact in the cross section will come from the various $(1,l)$ KK $Z$ bosons instead of higher $n$ models.  Thus we will focus on the strongly coupled $Z^{(1,l)}$ bosons.

\section{Results}
	
\begin{figure}
		\resizebox{0.9\columnwidth}{!}{ \includegraphics{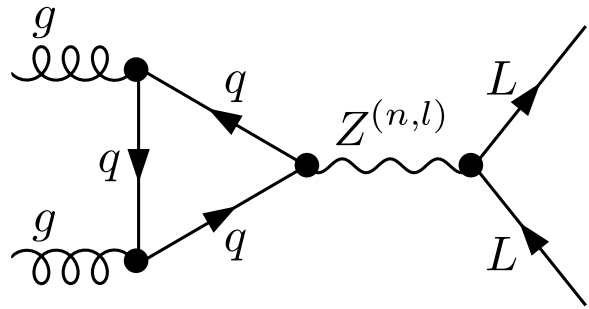} }
		\caption{Gluon fusion via Higgs or $Z$ boson to produce heavy charged lepton pairs.  In our analysis the quark loop can be any of the Standard Model quarks as well as the fourth generation quarks.  We assumed the $m_{t^\prime} = 500$ GeV and $m_{b^\prime} = 443$ GeV.}
		\label{fig.ggfusion}
	\end{figure}

	We consider $L\bar{L}$ pair production via gluon fusion through a quark loop to a Higgs, Z boson, or $Z_{nl}$, see fig.\ \ref{fig.ggfusion}.  We assumed the Higg's mass to be $120$ GeV.  Although, $kr_c$ was held fixed at $11$, $kR$ was varied from one to three.  We also chose the mass of $Z^{(1,1)}$ to be 2 TeV as Davoudiasl and Rizzo \cite{RS6} did (so we could compare our results more easily).
	
	We used FeynArts and FormCalc \cite{FeynArts, FormCalc} to generate the matrix element along with LoopTools, FORM, and LHAPDF \cite{FormCalc, FORM, LHAPDF} to complete the momentum integrals.
	
\begin{figure}
		\resizebox{1\columnwidth}{!}{ \includegraphics{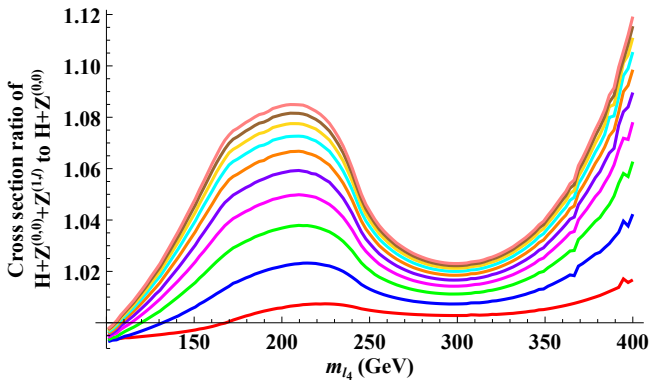} }
		\caption{The ratio of the cross-section including the $Z_{nl}$ contributions over the cross-section of only the SM $Z$ and Higgs.  $m_{l_4}$ is the mass of the charged lepton.  The bottom red line considers only the $Z_{11}$ boson, while the higher lines include contributions of higher $Z_{np}$ modes.}
		\label{fig.hz500}
	\end{figure}

	Remember that the $n=1$ modes dominate over higher $n$ modes.  As $l$ increases the $Z_{1,l}$ contribution decreases, but not as quickly as we expected.  This means that more modes need to be considered to have an accurate understanding of the cross-section.  
	The ratio of the process including the $Z_{nl}$ contributions to the Standard Model $Z$-boson and Higgs over just the Standard Model processes are shown in fig.\ \ref{fig.hz500}.
	
	However, as $kR$ increases the contribution of the higher $Z_{nl}$ modes to the cross-section also increases.  This is because the masses of the $Z_{nl}$ bosons become closer together.
	
	Of particular interest are the decay modes of the $L$.  If it is similar to its lighter cousins, the muon and the tau, then we expect to see a large decay mode $L\rightarrow N\ell\bar{\nu_\ell}$ where $\ell$ and $\bar{\nu_\ell}$ are a SM lepton and its associated anti-neutrino.  If the lepton comes from a $W$, then the neutrino is determined and the signature should be relatively straightforward to detect.
	
\section{Conclusions}
	
	The recent RS6 model with an added fourth generation has a clear $N+W$ signature.  Interestingly, the $Z$ boson excitations in the flat dimension (larger $l$) contribute more to the cross section than the higher $Z$ excitations in the warped dimension (larger $n$).

\vspace{1cm}
\begin{acknowledgments}
	We are grateful to Marc Sher for fruitful discussions.  Our work has been funded by St.\ Mary's College of Maryland.
\end{acknowledgments}

\end{document}